\documentclass[3p,times,twocolumn]{elsarticle}
 \biboptions{comma,sort&compress}
 
\usepackage{graphicx}
\usepackage{amsmath}
\usepackage{here}
\usepackage{xcolor}
\usepackage{ecrc}

\volume{00}

\firstpage{1}

\journalname{Nuclear and Particle Physics Proceedings}
\runauth{L. Abreu and F. J. Llanes-Estrada}

\jid{nppp}
\jnltitlelogo{Nuclear and Particle Physics Proceedings}

\usepackage{amssymb}

\usepackage[figuresright]{rotating}

\begin{document}

\begin{frontmatter}

\title{
%
Triangle singularities waning in Heavy Ion Collisions\, $^*$\\
the case of $Y(4260)\to \pi (\pi J/\psi)$} 
 
 \cortext[cor0]{Communication to QCD21 Montpellier - FR (held online). }

 \author[Bahia]{Luciano M. Abreu
 \corref{cor1} 
 }
 \author[UCM]{Felipe J. Llanes-Estrada
 \corref{cor2} 
 }

   \address[Bahia]{Instituto de F\'isica, Universidade Federal da Bahia, Salvador, Bahia, 40170-115, Brazil}
   \address[UCM]{Dept. F\'isica Te\'orica and IPARCOS, Universidad Complutense, Madrid, 28040, Spain}

\pagestyle{myheadings}
\markright{ }
\begin{abstract}
\noindent
We have recently observed that hadron triangle singularities, that can mock new exotic hadrons, can be significanttly
suppressed in relativistic heavy ion collisions, provided two conditions are met: these are, first, that the fireball lives long enough so that the triangle process has enough time to complete in the Norton-Coleman classical sense, and second, that the mass and/or width of the particles in the triangle diagram are sufficiently modified from their vacuum values. Here we add a very interesting example to the canon, which is $Y(4260) \to D_1 D \to \pi D^* D \to \pi + J/\psi \pi$. 
This reaction has been proposed as a mechanism to explain the appearance of $Z_c(3900)$ in the $J/\psi \pi$ spectrum. 
If the two muons and two pions reconstructing the initial-state $Y$ can be isolated from the combinatorial background, then the mechanism can provide a spectroscopy test: presence of $Y(4260)$ but absence of $Z_c(3900)$ would be more indicative of 
such triangle mechanism, while presence of both would rather point out to $Z_c$ being an exotic hadron.

\end{abstract}
\end{frontmatter}
\section{Triangle singularities or new hadrons?}

\begin{figure}[h]
\includegraphics[width=0.7\columnwidth]{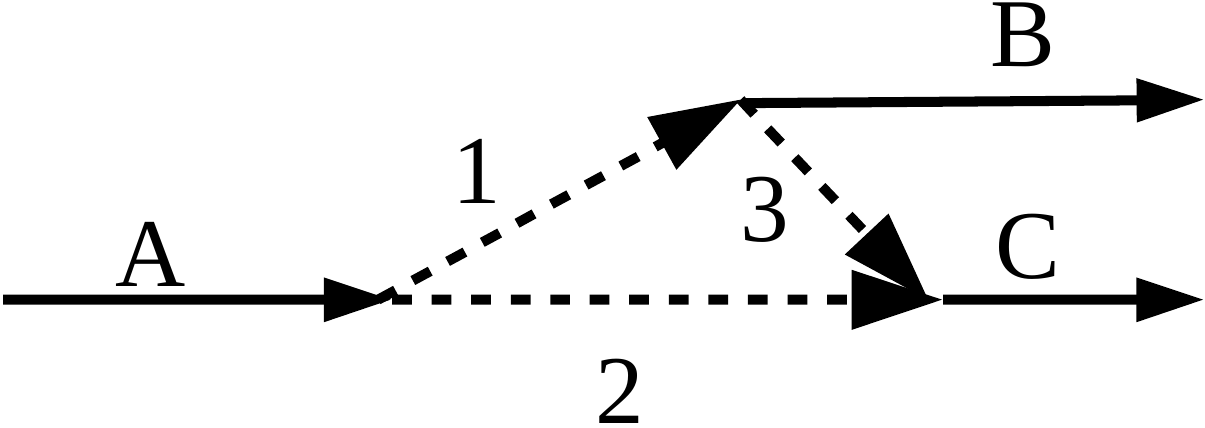}
\caption{\label{fig:triangle} A triangle loop can yield a singularity, according to the Coleman-Norton theorem,
when the kinematics of the external systems A, B, C are such as to (1) allow the internal particles 1, 2, 3 to
go on-shell, and (2) these three are collinear and their classical velocities allow 3 (emitted from 1) 
to catch up with 2.}
\end{figure}

Traditionally, hadron resonances were found in two-body (or sequential two-body) decays, such as 
$\tau\to \nu_\tau + \pi\pi$ where the neutrino escapes without rescattering and the pion pair strongly resonates in the $\rho$-channel. But in the last two decades, the extraordinary resolving power of modern detectors have yielded numerous sightings of possible new hadrons in more complicated decays involving several strongly interacting particles in the final state.

When a bump in a subset of a multihadron spectrum is observed, the possibility of a new hadron having been discovered is weighed against many other mechanisms that may enhance the corresponding cross-section, such as a threshold cusp, a Deck effect, or, in the case that occupies us here, a triangle singularity~\cite{Guo:2020oqk}. Such singularities (see fig.~\ref{fig:triangle}) arise when, in the transition $A\to BC$, an internal triangle graph with the particles 1, 2, 3, can 
occur with all of them simultaneously on-shell and when the reactions $A\to 1,2$; $1\to B,3$; $2,3\to C$ can all happen classically (Coleman-Norton theorem). 

Often, all that is needed to distinguish a true resonance from a triangle singularity is a different reaction. For example, the CLAS collaboration is trying to detect the claimed LHCb charmed pentaquark, seen in 
\begin{equation}
\Lambda^0_b \to K^- J/\psi p\ ,
\end{equation}
that could proceed with an intermediate triangle mechanism with the particles~\cite{Mikhasenko:2015vca} $(D_s^*D^*\Sigma_c)$ running in the loop,
in the alternative, triangle-free reaction
\begin{equation}
\gamma/\gamma^* p \to J/\psi p\ .
\end{equation}
If the supposed pentaquark does not appear in this production channel, the hypothesis that a triangle mechanism is the culprit will be very much favored.

An alternative is to study the same reaction of interest, but in a different setup, such as that provided by the medium in Relativistic Heavy Ion Collisions.
We have recently observed~\cite{Abreu:2020jsl} that a computation of the  triangle diagram in thermal field theory
(strictly valid for an infinite, stationary hadron gas),
\begin{eqnarray}    
\label{loop-finiteT} 
I_{\triangleleft} & \simeq &   \frac{1}{2} \int \frac{d^3 q}{(2 \pi)^3} \frac{1}{8 E_1 E_2 E_3} 
\frac{1}{\left(E_A - \tilde E_1 -\tilde E_2 \right)} \nonumber \\
& \times &   \frac{1}{\left( E_C - \tilde E_2 - \tilde E_3 \right)}   \frac{1}{\left( E_B - \tilde E_1 + \tilde E_3 \right)}  \nonumber \\   \nonumber
& \times & \left\{ \left[ 1 + 2n_\beta(\tilde E_2) \right] \left(\!  -E_B \!+\!  \tilde E_1\! -\! \tilde E_3\! \right) \right. 
\nonumber \\ &\phantom{\{}  & +
\left[ 1+2n_\beta(E_A- \tilde E_1 ) \right] \left(\! E_C \! -\!  \tilde E_2\! -\!  \tilde E_3\! \right)  \nonumber \\ &\phantom{\{}  & + \left.
   \left[ 1+2n_\beta (\tilde E_3-E_C) \right] 
\left(\!E_A\! -\! \tilde E_1\! -\! \tilde E_2\! \right) \right\}
\end{eqnarray}
can affect the kinematic accident causing its singular nature, softening the production line-shape of the final state. 
We found this to be due to the medium affecting the mass and width ($\tilde E_\alpha := E_\alpha-i\frac{\Gamma_\alpha}{2}$) of the particles in the loop.

\begin{figure}[h]
\centerline{\includegraphics[width=1.1\columnwidth]{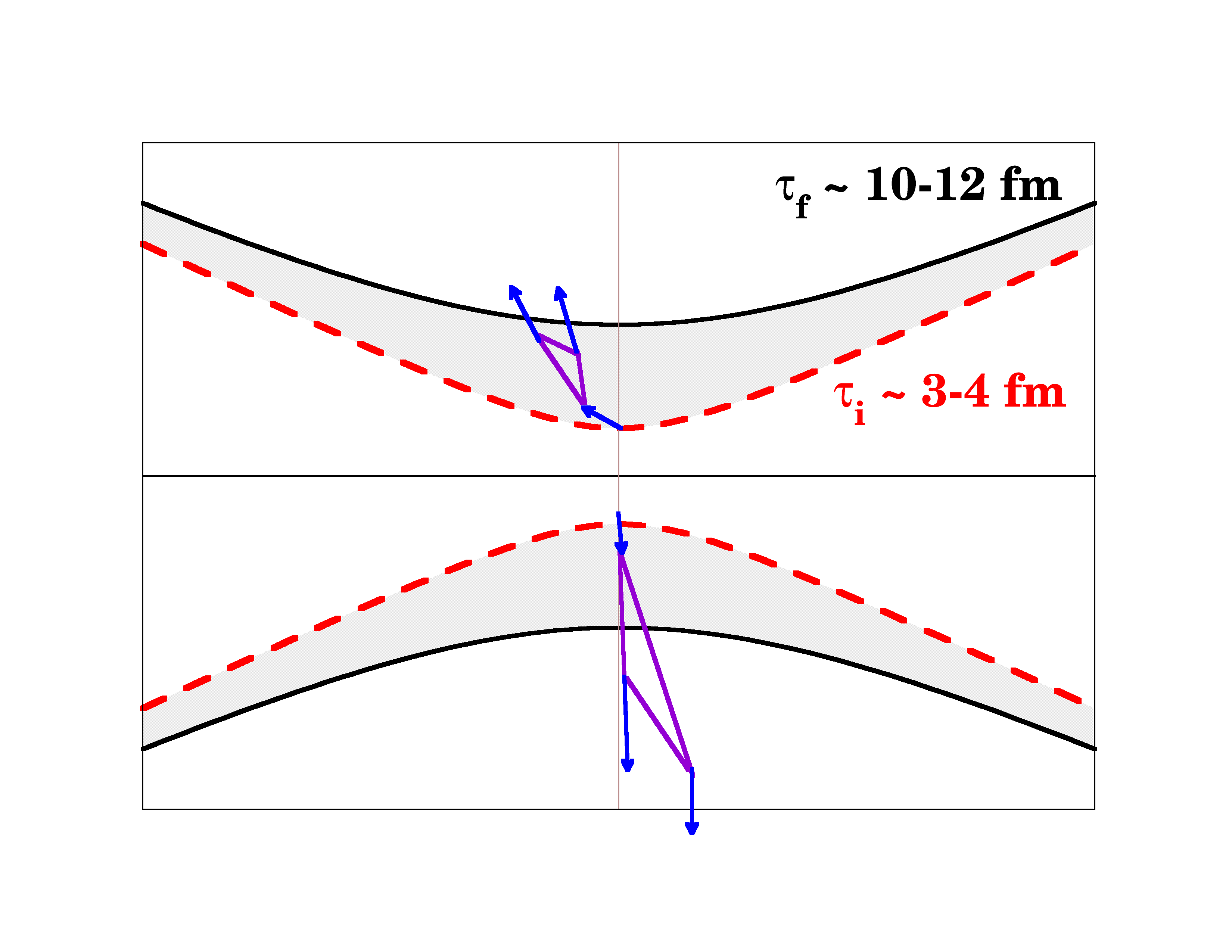}}
\caption{\label{fig:RHIC} Minkowski spacetime diagram showing hypersurfaces of equal Bjorken expansion time ($\tau=\sqrt{t^2-z^2}$ hyperbolae),
the duration and span of the hadron gas (shaded area), and two triangle diagrams. The top one completes in the lifetime of the collision, and thus has a significant chance of being affected by the medium. The bottom one does not close until well after the hadron gas has frozen out, because the characteristic triangle time is too long: therefore any triangle singularity will not be so strongly affected by the medium.}
\end{figure}

In a Relativistic Heavy Ion Collision, however, the medium expansion leads to eventual freeze out, so that the 
triangle reaction can complete only if its characteristic time is short enough compared to $O(10)$ fm, as shown in figure~\ref{fig:RHIC}. 

If these two conditions are met (the classical-kinematics time of flight of particle 3 is short, which allows to close the diagram soon enough, and one or more of the particles are severely affected by the medium), the triangle diagram softens and the cusp in the spectrum can disappear.
For example, the intensity of the peak is proportional~\cite{Debastiani:2018xoi} to the logarithm of the partial width  $\Gamma_{1\to B+3}$, that is, $I_{\triangleleft}\arrowvert_{\rm max} \propto \log (\Gamma_1) $; if this width is affected by the medium, the intensity of the triangle cusp readily reacts (this width also controls the classical time span in which the triangle diagram is active, see subsection~\ref{subsec:notime} below).

 Thus, if the mentioned conditions are met in a suspected triangle reaction, but the corresponding cusp enhancement is still seen in hadron collisions, one is lead to suspect that such triangle mechanism is not at play and that the given cross-section peak is possibly a new resonance. On the other hand, a disappearance of the spectral structure in heavy ion collisions can point out to a triangle singularity that is being erased, instead of a new hadron. Thus, experiments at RHIC or the LHC can distinguish between both types of data features. 

\section{A working example:\\ $Z_c(3900)$ in the decay of $Y(4260)$}

The $Y(4260)$ discovered by Babar in the $J/\psi\pi\pi$ final state~\cite{BaBar:2005hhc}, that could correspond to a charmonium $\psi(4260)$ since its $J^{PC}=1^{--}$ nonexotic quantum numbers allow it to be at least partly a pure $c\bar{c}$ meson~\cite{Llanes-Estrada:2005qvr}, has nonetheless certain nonconventional properties, such as appearing in a dip of $\sigma_{e^-e^+\to X}$ instead of a maximum, that have made it a focuse of intense exotic-hadron studies. 

Even more surprisingly, it has been found~\cite{Belle:2013yex,BESIII:2013ris} that a charge-exotic candidate $Z_c(3900)\to J/\psi \pi^\pm$ seems to populate the decay  $Y(4260)\to \pi^\mp Z_c$, so that two possibly exotic hadrons appear in the same reaction.

An alternative proposal is that the $Z_c$ signal supposedly seen in this reaction is a kinematic effect, see for example~\cite{Liu:2013vfa,Szczepaniak:2015eza} (or section 6.1.1 of~\cite{Guo:2020oqk} for a review discussion). 
This is due to the possible kinematic accident of the $(D^*_1DD^*)$ triangle in the loop satisfying the Coleman-Norton theorem, thus yielding a peaked spectrum, as in the reaction
\begin{eqnarray}
Y(4260)\ (A) \to \underbrace{D_1\ (1)}_{\to \pi\ (B) D^*\ (3)} D\ (2) \nonumber \\ 
D^*\ (3) D\ (2) \to (J/\psi \pi)\ (C)  \ .
\label{reaction}
\end{eqnarray}
Indeed, it was noted early on~\cite{Close:2005iz} 
that the $\psi(4260)$ sits just below the s-wave threshold for production of a $c\bar{c}$ pair that separate on a pair of $S$-wave mesons; this is the $D_1 (1^{+}) D (0^-)$ threshold
that is therefore expected to influence reactions involving $\psi(4260)$.

Likewise, the position of this reaction in the plane $(M_{J/\psi\pi\pi},M_{J/\psi \pi^\pm})$ is only 30 MeV away from the triangle singularity~\cite{Guo:2020oqk}. Therefore, the possibility that this supposed resonance is really a structure caused by the $t$-channel ``force'' induced by that $D_1D^*$ exchange cannot be discounted.

This is a reaction where heavy-ion collisions can eventually help discern the correct interpretation of the data; 
this is because the involved $D$-family mesons are sufficiently affected by the medium to yield an effect, and the $D_1$ is sufficiently short-lived that the triangle diagram is completed in the requisite short time.

Indeed, because the time of flight as described shortly in Eq.~(\ref{timeofF}) is $\tau_A< 10$ fm/c (choosing an appropriate value of $m_C$; with $\Gamma_{D_1}= 31$ MeV and also 
$\Gamma_{Y(4260)} > 54$ MeV), the triangle process has sufficient time to complete in a relativistic heavy ion collision.

\begin{figure}
\includegraphics[width=\columnwidth]{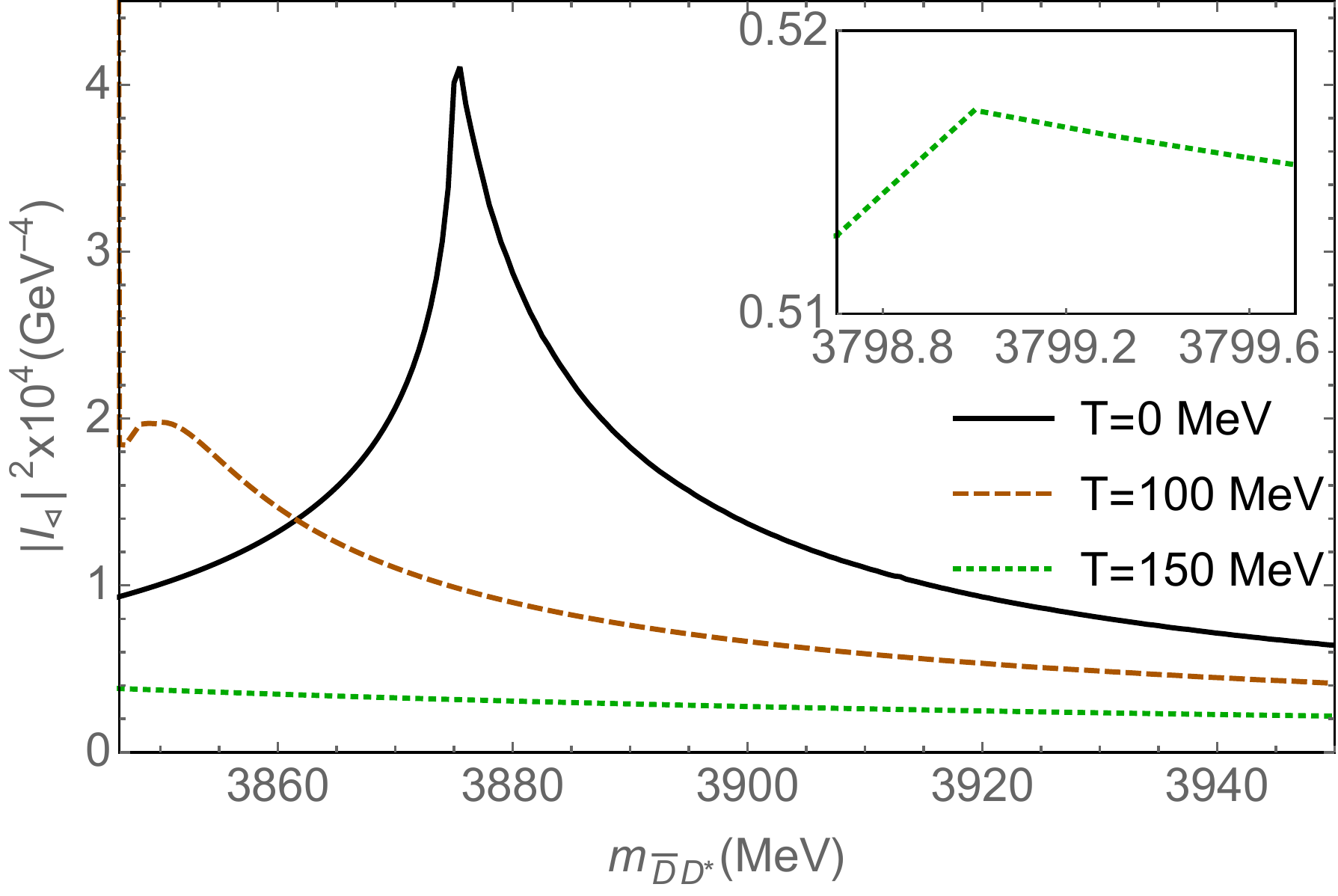}
\caption{\label{fig:Zc} 
At zero temperature (solid line, black online), the triangle diagram produces a clear peaking structure that can mock a resonance. When the temperature is increased, modifying the in-medio masses and widths of the $D$ mesons in the loop according to Table~\ref{thermalmasses}, the singularity drops; at $T=150$ MeV, near the maximum temperature of a hadron gas, the singularity has basically disappeared. Thus, heavy ion collisions can help decide whether the charged charmonium $Z_c(3900)$ is or not an intrinsic state.}
\end{figure}

Figure~\ref{fig:Zc} shows that we reproduce the reported peaking singularity in the loop of fig.~\ref{fig:triangle}  and that the effect of the medium on this singularity is marked, making it all but disappear at the temperature of 150 MeV just below the transition to the quark-gluon plasma. 

\begin{center}
\begin{table}[h]
\caption{Meson thermal masses and widths extracted from Ref.~\cite{Montana:2020vjg} used as input in the squared scalar triangle loop integral, defined in Eq. (\ref{loop-finiteT}), for the reaction in Eq.~(\ref{reaction}). 
We assumed $m_\pi$  constant (see Ref.~\cite{Llanes-Estrada:2005qvr} for a detailed discussion). The mass of the initial state is fixed at $m_A = 4.285$ GeV. We are not currently aware of a computation of the $D_1$ mass and width, so we use a rough estimate based on \cite{Montana:2020vjg} for the excited states. All quantities are in GeV.
\label{thermalmasses}  } 
\begin{center}
\begin{tabular}{c|c|c|c}
\hline
\hline
T & $ 0$  & $ 0.1$ & $ 0.15$   \\ 
\hline
$m_{\pi} $        & 0.1396  & 0.1396 & 0.1396  \\
$m_{D^{\ast }} $  & 2.0103  & 1.994  & 1.872 \\
$m_{D^{0}} $      & 1.8648  & 1.856  & 1.827  \\
$m_{D_1} $        & 2.4181  & 2.4104 & 2.4071 \\
$\Gamma_{D^{\ast}}$ & $83.4 \times 10^{-6}$ & 0.0079 & 0.071 
\\
$\Gamma_{D_1} $   & 31.3    & 35     & 40 
\\
\hline
\hline
\end{tabular}
\end{center}
\end{table}
\end{center}

\section{Two examples where the method is indecisive}

There are many examples of triangle diagrams where Relativistic Heavy Ion Collisions can be useful~\cite{Abreu:2020jsl}, among them $N^* \to K \Lambda(1405)$ with $(K^*\Sigma\pi)$ in the triangle; production of $\pi^0\pi^+\pi^-$ or $\pi^0\pi^0\eta$ with $(K^*K^-K^+)$ in the loop, that we will consider in subsection~\ref{subsec:KKK} below; or
$pp\to \pi^+ d$ with $(\Delta(1232)pn)$ which is of interest for the claimed deuteron excited state.
However, there are two clearly identified circumstances whence the medium does not affect the triangle diagram in a meaningful way, so that detecting the presumed particle following the RHIC medium does not help with  its spectroscopic classification. We quickly examine these two cases by means of two simple examples.

\subsection{ $c\bar{c}\to \pi^- (\pi\pi J/\psi)$ with $D^{*-}D^{*0}D^0$ in the triangle: no time to close the triangle during the collision.}\label{subsec:notime}

This reaction would in principle be affected by the medium, as we have shown in~\cite{Abreu:2020jsl}, but it does not have time to complete. 

The key quantity to compute is the time of flight of particle 2, or equivalently, the sum of the times flown by particle 1 before its decay and particle 3 thereafter, that is; the time since the decay of $A$ until the coalescence of 2 and 3. This can be given, in the rest frame of particle $A$ as 
\begin{equation} \label{timeofF}
\tau_A =\frac{\gamma(\beta_1)}{\Gamma_1} \frac{\beta_3-\beta_1}{\beta_3-\beta_2}
\end{equation}
in terms of the various velocities $\beta_i$, the Lorentz dilation factor $\gamma$ of particle 1, and its (partial) width $\Gamma_1$ for the decay $1\to B+3$.

With the kinematics of this reaction, all particles in the loop being on-shell as per the Coleman-Norton theorem, we estimate $\tau_A=2346 {\rm\ fm\ } \gg 10 {\rm\ fm}$ indicating that if the triangle is present in a RHIC event, it completes its course only well after the system has undergone freeze out; therefore its absence or presence is not informative.

\subsection{Including {\it versus} disregarding the effect of the medium on the loop particles}
\label{subsec:KKK}

We now consider an example  with a  $K^{*}  \overline{K} K$ triangle diagram,  that appears in the putative $a_1(1420)$ production by COMPASS in the $P-$wave $\pi f_0(980)$ channel of the $\pi p \rightarrow \pi \pi \pi p$ reaction~\cite{Adolph:2015pws}. 

The process relevant to the $K^{\ast} K$ threshold energy regime is  
\begin{eqnarray} \label{reaccion2}
{\rm A}\! \rightarrow\! K^{*+}(1)  K^{-}(2)  K^+ (3)\! \rightarrow \!
\pi^0 ({\rm B}) 
(\pi^+ \pi^-) / (\pi^0 \eta)
({\rm C}) .
\end{eqnarray}

\begin{figure}[h]
\centering
\includegraphics[width=\columnwidth]{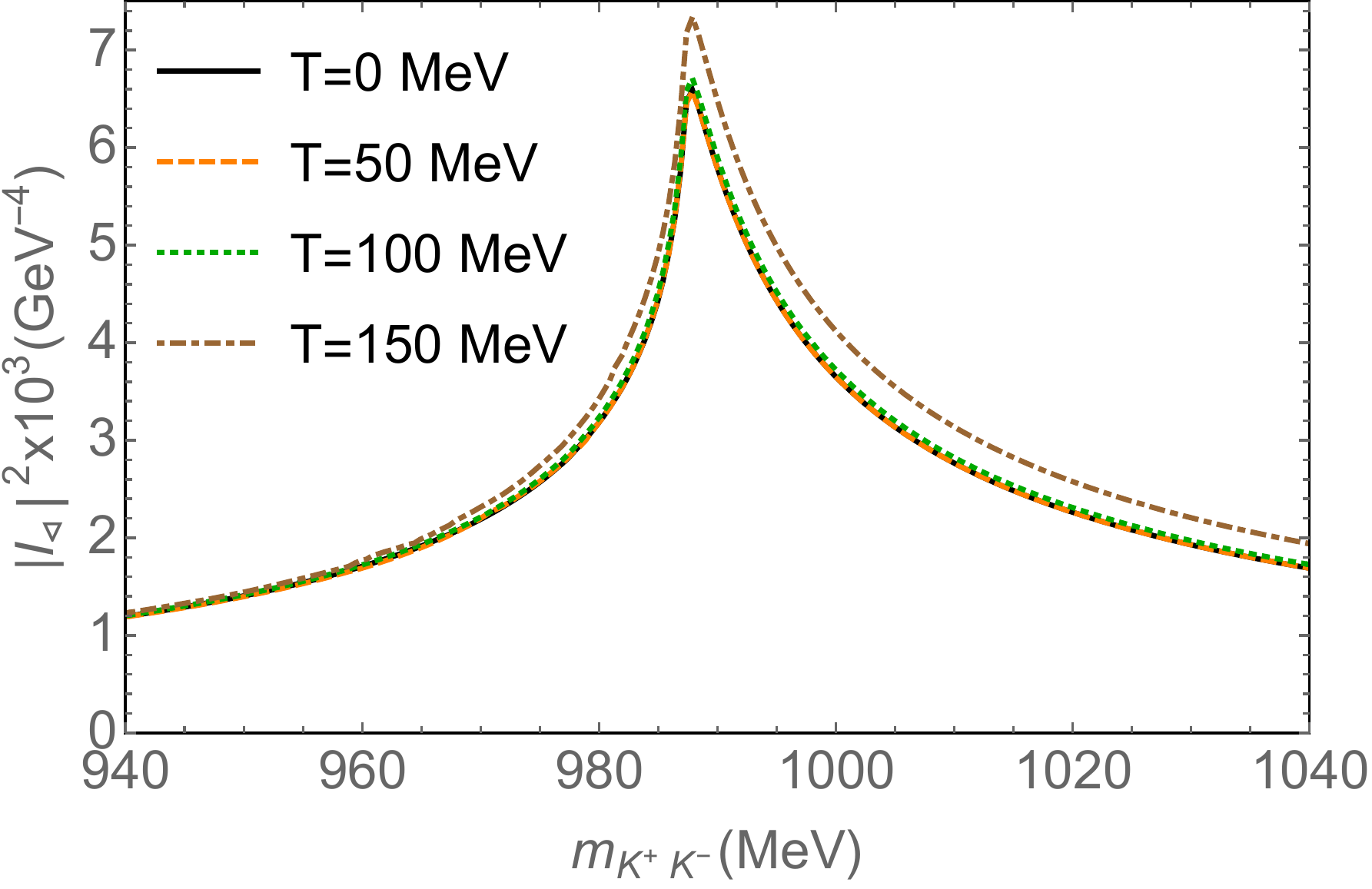} \\
\includegraphics[width=\columnwidth]{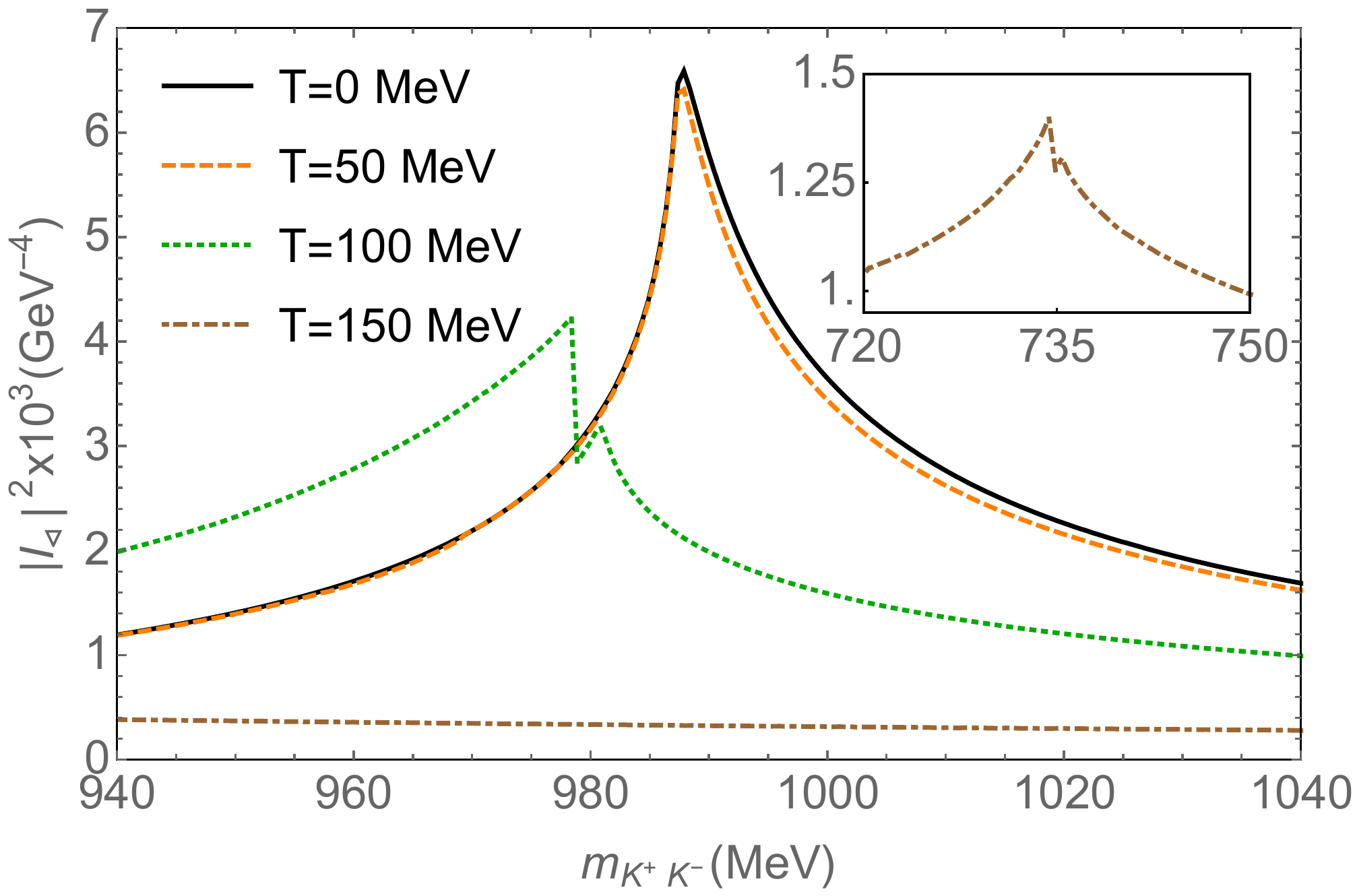} 
\caption{Squared, scalar, $K^*KK$ triangle loop integral $(|I_\triangleleft|^2)$, as a function of the $ K^+ K^-$ invariant mass, but with fixed 1420 MeV $K^{*+} K^- $ mass, for various $T$. 
{\bf Top}: temperature included only in the loop variable $q_0$, but $M_i$, $\Gamma_i$  fixed at their vacuum values (except for a regulating $\Gamma _2 = 0.5$ MeV to avoid numerical instability, so the $T=0$ peak is underestimated). 
{\bf Bottom}: thermal corrections to the $(1,2,3)$ meson masses and widths are included. The singularity melts with $T$. 
\emph{(Reprinted from \cite{Abreu:2020jsl} under STM permissions guidelines)}
\label{fig:3K}}
\end{figure}

The inclusion of thermal effects in the triangle loop integral alone 
barely shifts or weakens the triangle singularity, as visible in the top panel of Fig.~\ref{fig:3K}.
On the contrary, thermal corrections to the masses and widths do affect  the triangle singularity  (bottom plot of Fig.~\ref{fig:3K}). 

At zero or small $T$ the narrow peak of the triangle singularity is clearly visible.
It becomes less pronounced, while shifting to smaller energy, accompanying the motion 
of the $K^{+} K^- $ threshold. The threshold and triangle singularities are also seen to 
separate into two nearby structures.

We have found the features of this example in several others. In particular, the 
small effect of the temperature on the triangle itself, with sizeable modifications only 
when the in-medio propagators are used, seems general. Because pion masses are barely modified at finite temperature (in fact, it is not clear from the literature whether the physical pion pole mass increases or decreases in medio), triangles formed with pions in the loop are not expected to be modified by the medium. 

\newpage
\section{Conclusion}

The study of spectroscopy in heavy-ion collisions~\cite{Cho:2011ew} is generally less competitive than in other type of experiments, because of the large combinatorial backgrounds due to large particle multiplicities, that make difficult to match the correct ones coming from the same decays. However, in this work we have identified a niche where collaborations such as STAR or ALICE can make a contribution, in searching for resonances that are doubted to arise from triangle singularities. If the two mentioned requirements (short characteristic time and modified masses or widths of the loop particles) are satisfied, which is easy to check on a case by case basis, the medium can serve as a spectroscopic filter and help distinguish actual hadrons from singular production mechanisms with a triangle loop.

\newpage 
\section*{Acknowledgment}
This work was partially supported by the following grants: 
spanish MINECO FPA2016-75654-C2-1-P, MICINN PID2019-108655GB-I00 and -106080GB-C21; 
EU’s 824093 (STRONG2020); and UCM’s 910309 group grants and IPARCOS, as well as
the Brazilian CNPq (contracts 308088/2017-4 and 400546/2016-7) and FAPESB (contract
INT0007/2016).  


\end{document}